\documentclass[doublespacing]{elsart}

\usepackage{latexsym}
\usepackage{amstext,amssymb,amsfonts}
\usepackage{epsfig}

\begin{document}

\begin{frontmatter}



\title{Anomalous electrical and frictionless flow conductance 
in complex networks}

\author[1]{Eduardo L\'{o}pez}
\author[2]{Shai Carmi}
\author[2,3]{Shlomo Havlin}
\author[3,4]{Sergey V. Buldyrev}
\author[3]{H. Eugene Stanley}
\address[1]{Theoretical Division, Los Alamos National Laboratory, Mail Stop B258, Los Alamos, NM 87545 USA}
\address[2]{Minerva Center \& Department of Physics, Bar-Ilan University, Ramat Gan, Israel}
\address[3]{Center for Polymer Studies, Boston University, Boston, MA 02215 USA}
\address[4]{Department of Physics, Yeshiva University, 500 West 185th Street, New York, NY 10033 USA}

\begin{abstract}
We study transport properties such as electrical and frictionless flow
conductance on
scale-free and Erd\H{o}s-R\'{e}nyi networks. We consider the
conductance $G$ between two arbitrarily chosen nodes
where each link has the same unit resistance. Our theoretical
analysis for scale-free networks predicts a broad range
of values of $G$, with a
power-law tail distribution $\Phi_{\rm SF}(G)\sim G^{-g_G}$, where
$g_G=2\lambda -1$, where $\lambda$ is the decay exponent for the scale-free network degree distribution.
We confirm our predictions by simulations of
scale-free networks solving the Kirchhoff equations for the conductance between
a pair of nodes. The power-law tail
in $\Phi_{\rm SF}(G)$ leads to large values of $G$, thereby
significantly improving the transport in scale-free networks, compared
to Erd\H{o}s-R\'{e}nyi networks where the tail of the conductivity
distribution decays exponentially. Based on a simple physical
``transport backbone'' picture we suggest that the conductances of
scale-free and Erd\H{o}s-R\'{e}nyi networks can be approximated by
$ck_Ak_B/(k_A+k_B)$ for any pair of nodes $A$ and $B$ with degrees $k_A$
and $k_B$. Thus, a single quantity $c$, which depends on the average
degree $\overline{k}$ of the network, characterizes transport on both
scale-free and Erd\H{o}s-R\'{e}nyi networks. We determine that $c$
tends to 1 for increasing $\overline{k}$, and it is larger for
scale-free networks.
We compare the electrical results with a model for frictionless transport,
where conductance is defined as the number of link-independent paths
between $A$ and $B$, and find that a similar picture holds.
The effects of distance on the value of conductance are considered for
both models, and some differences emerge. Finally, we use a recent
data set for the router level of the Internet and confirm that our
results are valid in this real-world example.

\end{abstract}

\begin{keyword}
Complex networks \sep
Transport \sep
Diffusion \sep
Conductance \sep
Scaling

\PACS 89.75.Hc \sep 05.60.Cd
\end{keyword}
\end{frontmatter}

\section{Introduction}

Transport in many random structures is ``anomalous,'' i.e.,
fundamentally different than that in regular space
\cite{havlin-ben-avraham,ben-avraham-havlin,bunde-havlin}.
The anomaly is due to the random substrate on which transport is
constrained to take place.
Random structures are found in many places in the real world, from oil
reservoirs to the Internet, making the study of anomalous transport
properties a far-reaching field.  In this problem, it is paramount to
relate the structural properties of the medium with the transport
properties.

An important and recent example of random substrates is that of complex
networks.  Research on this topic has uncovered their importance for
real-world problems as diverse as the World Wide Web and the Internet to
cellular networks and sexual-partner networks \cite{rev-Albert}.

\begin{figure}[t]
\begin{center}
\epsfig{file=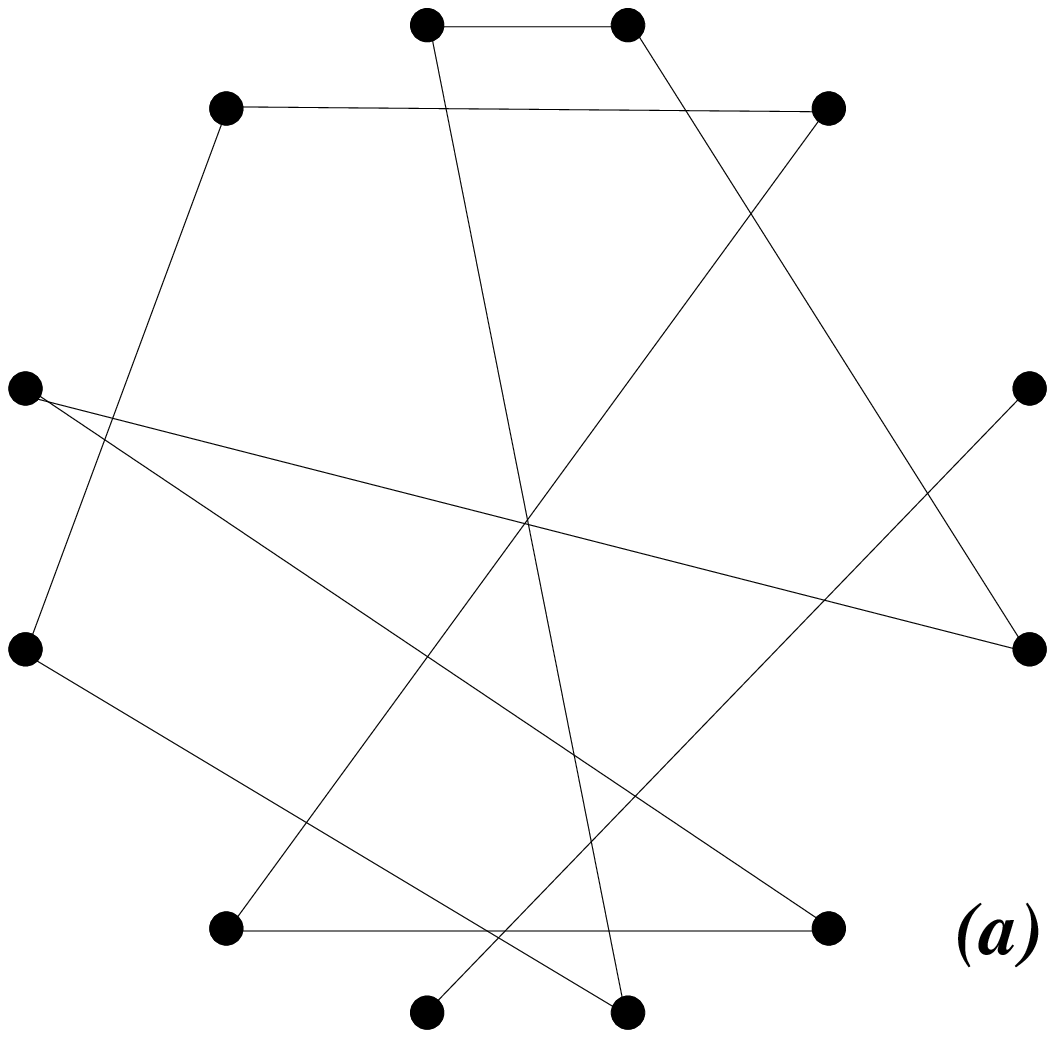,height=3.6cm,width=3.8cm}
\hspace{0.5cm}
\epsfig{file=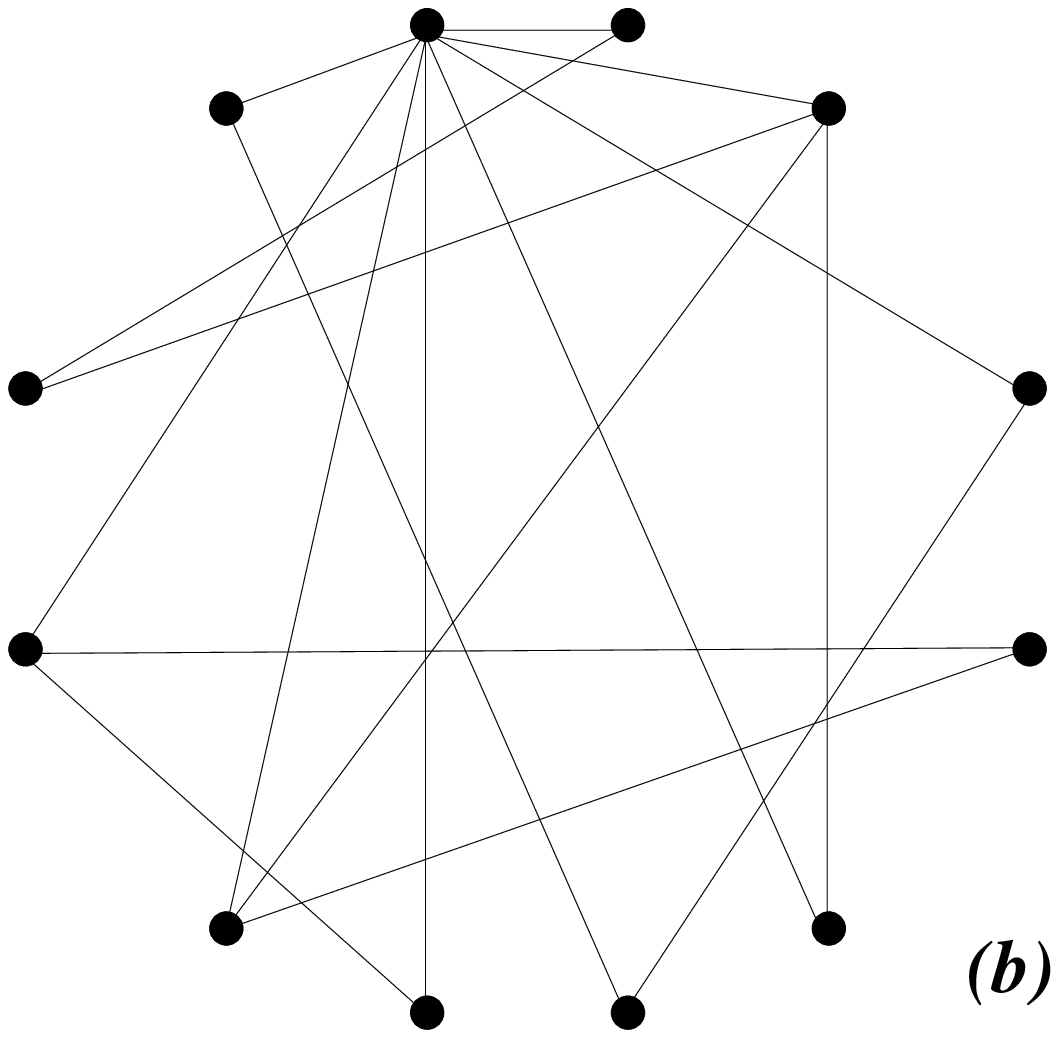,height=3.6cm,width=3.8cm}
\end{center}
\caption{(a) Schematic of an Erd\H{o}s-R\'{e}nyi network of $N=12$ and
$p=1/6$. Note that in this example ten nodes
have $k=2$ connections, and two nodes have $k=1$ connections.
This illustrates the fact that for Erd\H{o}s-R\'{e}nyi networks,
the range of values of degree is very narrow, typically close to
$\overline{k}$.
(b) Schematic of a scale-free network of $N=12$, $k_{\rm min}=2$ and
$\lambda\approx 2$. We note the presence of a hub with $k_{\rm max}=8$
which is connected to many of the other links of the network.}
\label{ER_diag}
\end{figure}

Two distinct models describe the two limiting cases for the structure
of the complex networks. The first of these is the classic Erd\H{o}s-R\'{e}nyi
model of random networks \cite{ER}, for which sites are connected with a
link with probability $p$ and disconnected (no link) with probability
$1-p$ (see Fig.~\ref{ER_diag}).
In this case the degree distribution $P(k)$,
the probability of a node to have $k$ connections, is a Poisson
\begin{equation}
P(k)\sim \frac{\left(\overline{k}\right)^k e^{-\overline{k}}}{k!},
\label{Poisson}
\end{equation}
where $\overline{k}\equiv\sum_{k=1}^{\infty} kP(k)$ is the average
degree of the network.  Mathematicians discovered critical
phenomena through this model. For instance, just as in percolation
on lattices, there is a critical value $p=p_c$ above which the
largest connected component of the network has a mass that scales
with the system size $N$, but below $p_c$, there are only small
clusters of the order of $\log N$. Another characteristic of an
Erd\H{o}s-R\'{e}nyi network is its ``small-world'' property which
means that the average distance $d$ (or diameter) between all
pairs of nodes of the network scales as $\log N$ \cite{Bollobas}.
The other model, recently identified as the characterizing
topological structure of many real world systems, is the
Barab\'{a}si-Albert scale-free network and its extensions
\cite{scale-Barabasi,dyn-network,Simon}, characterized by a
scale-free degree distribution:
\begin{equation}
P(k)\sim k^{-\lambda}\qquad [k_{\rm min}\le k\le k_{\rm max}],
\label{degree}
\end{equation}
The cutoff value $k_{\rm min}$ represents the minimum allowed value of
$k$ on the network ($k_{\rm min}=2$ here), and $k_{\rm max}\equiv k_{\rm
min}N^{1/(\lambda-1)}$, the typical maximum degree of a network with $N$
nodes \cite{Cohen,netwcomment}.  The scale-free feature allows a network
to have some nodes with a large number of links (``hubs''), unlike the
case for the Erd\H{o}s-R\'{e}nyi model of random networks \cite{ER,Bollobas}.
Scale-free networks with $\lambda >3$ have $d\sim \log N$, while for
$2<\lambda <3$ they are ``ultra-small-world'' since the diameter
scales as $d\sim \log\log N$ \cite{rev-Albert,Cohen}.

Here we extend our recent study of transport in complex networks
\cite{Lopez-PRL,book-chap}.
We find that for scale-free networks with $\lambda\ge 2$, transport
properties characterized by conductance display a power-law tail
distribution that is related to the degree distribution $P(k)$.
The origin of this power-law tail is due to pairs of
nodes of high degree which have high conductance.
Thus, transport in scale-free networks is better because of the presence
of large degree nodes (hubs) that carry much of the traffic,
whereas Erd\H{o}s-R\'{e}nyi networks lack hubs and the transport
properties are controlled mainly by the
average degree $\overline{k}$ \cite{Bollobas,Grimmett-Kesten}.
Also, we present a simple physical picture of transport in
scale-free and Erd\H{o}s-R\'{e}nyi networks and test it through simulations.
Additionally, we study a form of frictionless transport, in which
transport is measured by the number of independent paths between source
and destination. These later results are similar to those in \cite{Lee}.
The results of our study are relevant to problems of transport in
scale-free networks, given that conductivity and diffusivity are related
by the Einstein relation \cite{havlin-ben-avraham,ben-avraham-havlin,bunde-havlin}.

The paper is structured as follows. Section~\ref{transport}
concentrates on the numerical calculation of the electrical
conductance of networks. In Sec.~\ref{theory} a simple physical
picture gives a theoretical explanation of the results.
Section~\ref{frictionless} deals with the number of
link-independent paths as a form of transport. In
Sec.~\ref{summary} we present the conclusions and summarize the
results in a coherent picture.

\section{Transport in complex networks}
\label{transport}

Most of the work done so far regarding complex networks has concentrated
on static topological properties or on models for their growth
\cite{rev-Albert,Cohen,dyn-network,Toroczkai}.  Transport features have
not been extensively studied with the exception of random walks on
specific complex networks \cite{Noh,Sood,Gallos}. Transport
properties are important because they contain information about network function
\cite{dyn-topology}.  Here we study the electrical conductance $G$
between two nodes $A$ and $B$ of Erd\H{o}s-R\'{e}nyi and scale-free
networks when a potential difference is imposed between them.  We assume
that all the links have equal resistances of unit value
\cite{comm-struc}.

To construct an Erd\H{o}s-R\'{e}nyi network, we begin with $N$ nodes and
connect each pair with probability $p$.
To generate a scale-free network with $N$ nodes, we use the Molloy-Reed
algorithm \cite{Molloy-Reed}, which allows for the construction of
random networks with arbitrary degree distribution.
We generate $k_i$
copies of each node $i$, where $k_i$ is a random number taken from a
distribution of the form
$P(k_i)\sim k_i^{-\lambda}$.
We then randomly pair these copies of the
nodes in order to construct the network, making sure that two
previously-linked nodes are not connected again, and also excluding
links of a node to itself \cite{fn1}.

We calculate the conductance $G$ of the network between two nodes $A$
and $B$ using the Kirchhoff method \cite{Kirchhoff}, where entering and
exiting potentials are fixed to $V_A=1$ and $V_B=0$.
We solve the set of $N-2$ linear equations
\begin{equation}
\sum_{j=1,j\neq i}^N\frac{V_j-V_i}{r_{ij}}=0, \qquad \forall i\neq A,B
\end{equation}
representing the conservation of current at the nodes.
The resistances $r_{ij}$ are 1 if nodes $i$ and $j$ are connected, and
infinite if $i$ and $j$ are not connected.
Finally, the total current $I\equiv G$ entering at node $A$ and
exiting at node $B$ is computed by adding the outgoing currents from $A$
to its nearest neighbors through $\sum_{j}(V_A-V_j)$, where $j$ runs
over the neighbors of $A$.

\begin{figure}[t]
\begin{center}
\epsfig{file=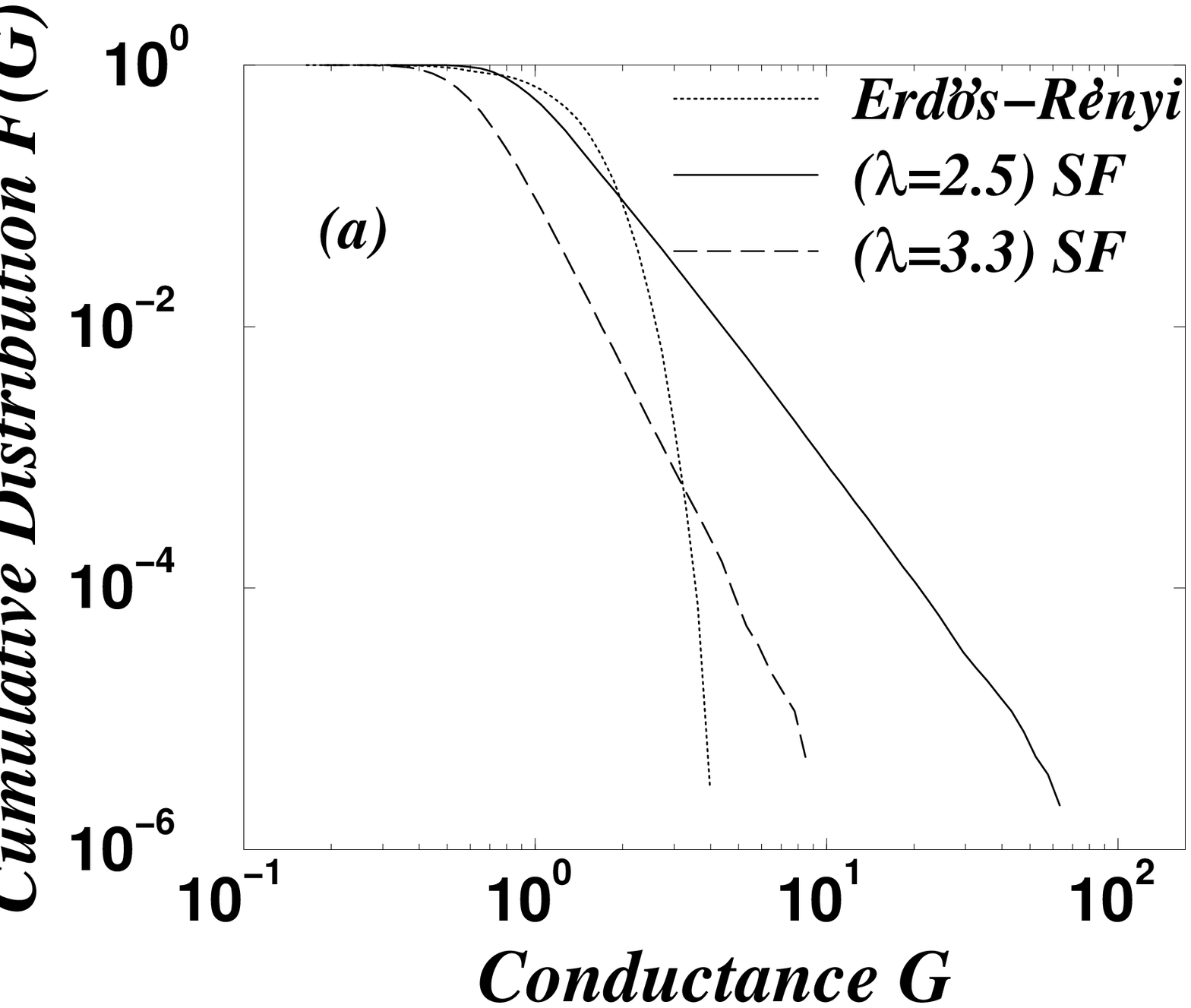,height=6cm,width=6cm}
\hspace{0.5cm}
\epsfig{file=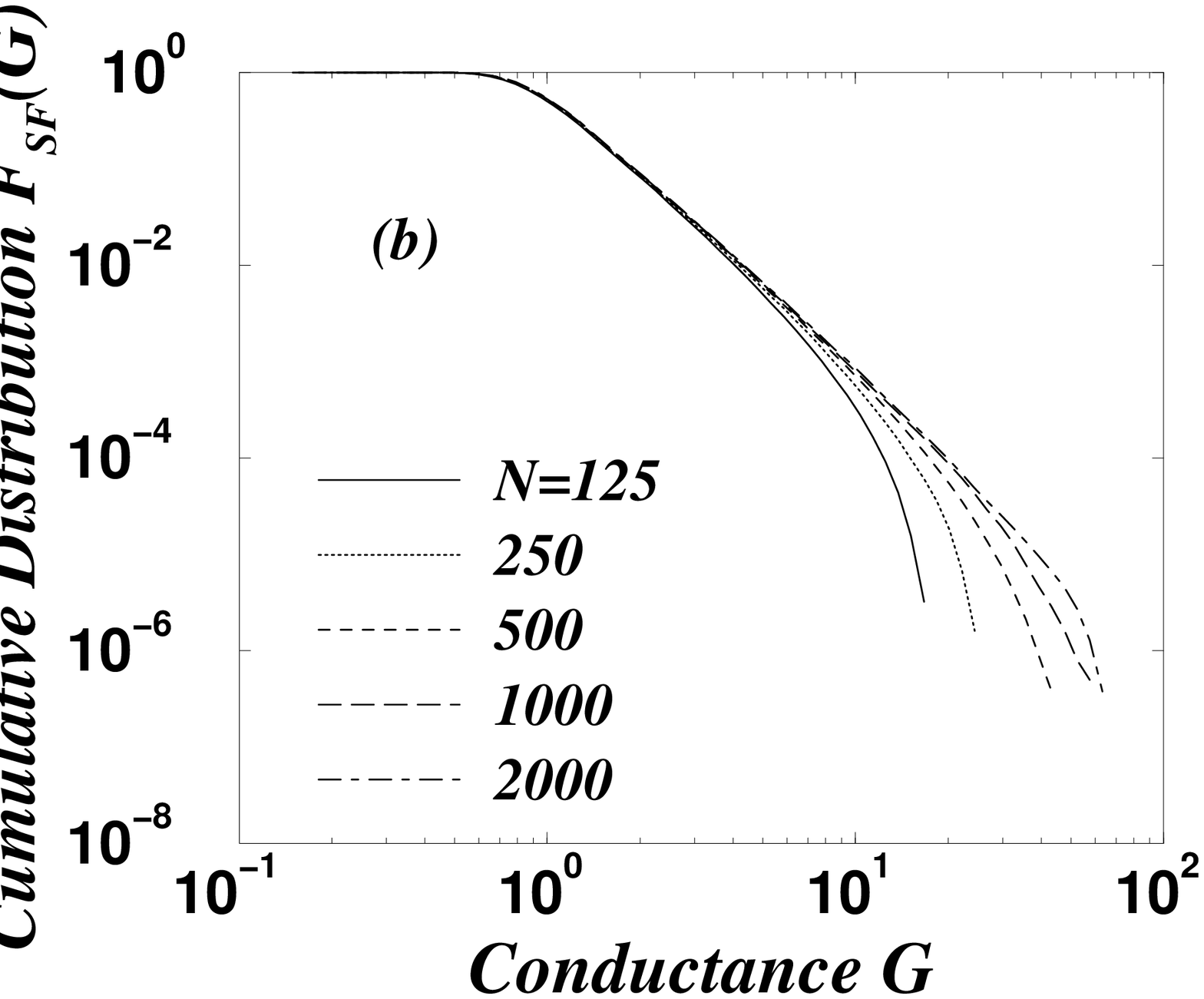,height=6.2cm,width=6cm}
\end{center}
\caption{(a) Comparison for networks with $N=8000$ nodes between the
  cumulative distribution functions of conductance for the Erd\H{o}s-R\'{e}nyi and the
  scale-free cases (with $\lambda=2.5$ and 3.3). Each curve represents
  the cumulative distribution $F(G)$ vs. $G$. The simulations have at
  least $10^6$ realizations.
  (b) Effect of system size on $F_{\rm
  SF}(G)$ vs. $G$ for the case $\lambda=2.5$. The cutoff value of the
  maximum conductance $G_{\rm max}$ progressively increases as $N$
  increases.  }
\label{FG_lamb2.5-3.3-ER_N8000}
\end{figure}

First, we analyze the probability density function (pdf) $\Phi(G)$ which
comes from $\Phi(G)dG$, the probability that two nodes on the network
have conductance between $G$ and $G+dG$.  To this end, we introduce the
cumulative distribution $F(G)\equiv\int_{G}^\infty \Phi(G')dG'$, shown
in Fig.~\ref{FG_lamb2.5-3.3-ER_N8000}(a) for the Erd\H{o}s-R\'{e}nyi and
scale-free ($\lambda=2.5$ and $\lambda=3.3$, with $k_{\rm min}=2$)
cases.  We use the notation $\Phi_{\rm SF}(G)$ and $F_{\rm SF}(G)$ for
scale-free, and $\Phi_{\rm ER}(G)$ and $F_{\rm ER}(G)$ for
Erd\H{o}s-R\'{e}nyi.  The function $F_{\rm SF}(G)$ for both
$\lambda=2.5$ and 3.3 exhibits a tail region well fit by the power law
\begin{equation}
F_{\rm SF}(G)\sim G^{-(g_G-1)},
\end{equation}
and the exponent $(g_G-1)$ increases with $\lambda$.  In contrast,
$F_{\rm ER}(G)$ decreases exponentially with $G$.

Increasing $N$ does not significantly change $F_{\rm SF}(G)$
(Fig.~\ref{FG_lamb2.5-3.3-ER_N8000}(b)) except for an increase in the
upper cutoff $G_{\rm max}$, where $G_{\rm max}$ is the typical maximum
conductance, corresponding to the value of $G$ at which $\Phi_{\rm
SF}(G)$ crosses over from a power law to a faster decay. We observe no
change of the exponent $g_G$ with $N$. The increase of $G_{\rm max}$
with $N$ implies that the average conductance $\overline{G}$ over all
pairs also increases slightly.

We next study the origin of the large values of $G$ in scale-free
networks and obtain an analytical relation between $\lambda$ and $g_G$.
Larger values of $G$ require the presence of many parallel paths, which
we hypothesize arise from the high degree nodes.  Thus, we expect that
if either of the degrees $k_A$ or $k_B$ of the entering and exiting
nodes is small (e.g. $k_A>k_B$), the conductance $G$ between $A$ and $B$
is small since there are at most $k$ different parallel branches coming
out of a node with degree $k$.  Thus, a small value of $k$ implies a
small number of possible parallel branches, and therefore a small value
of $G$.  To observe large $G$ values, it is therefore necessary that
both $k_A$ and $k_B$ be large.

\begin{figure}[t]
\begin{center}
\epsfig{file=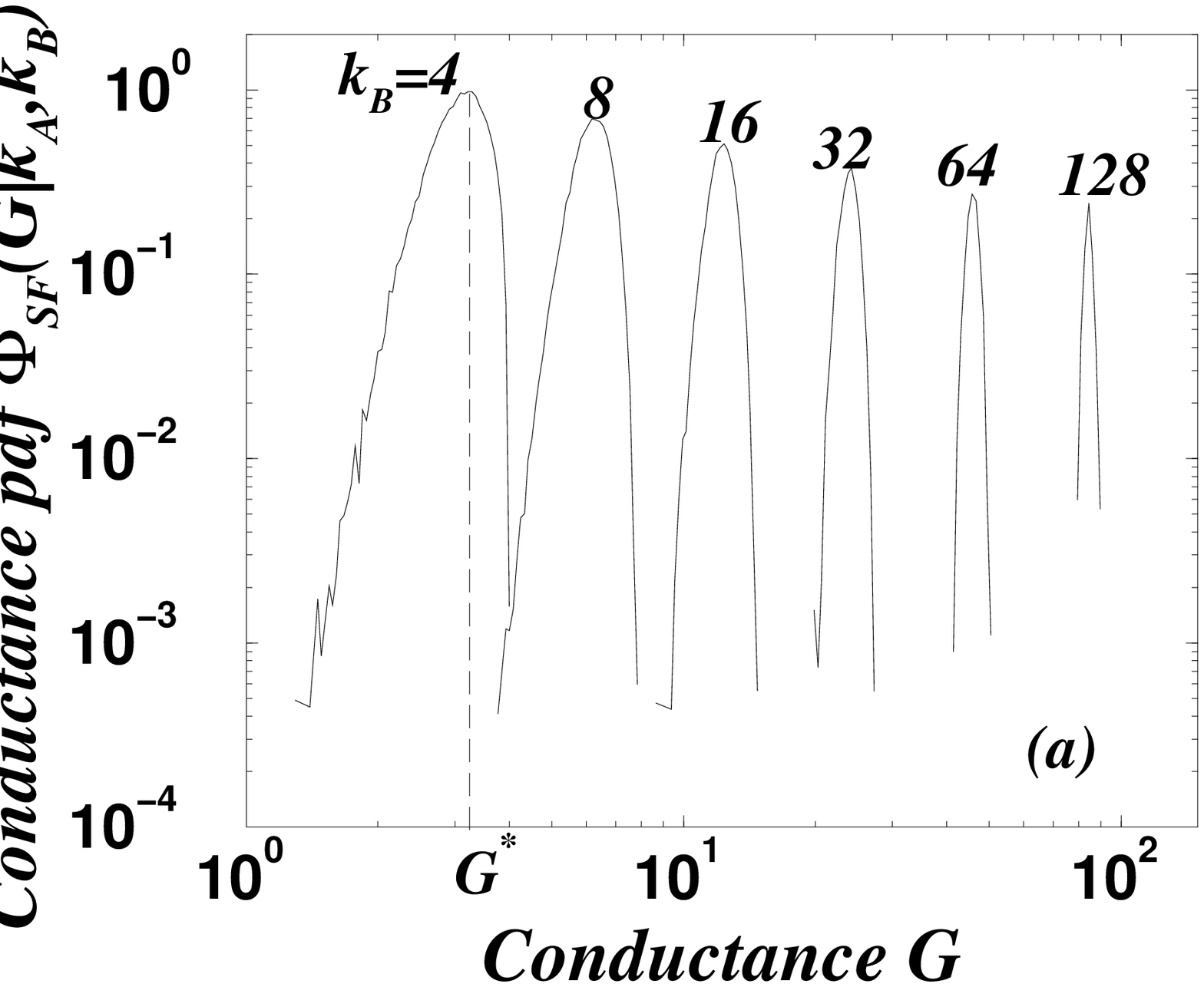,height=6.2cm,width=6cm}
\epsfig{file=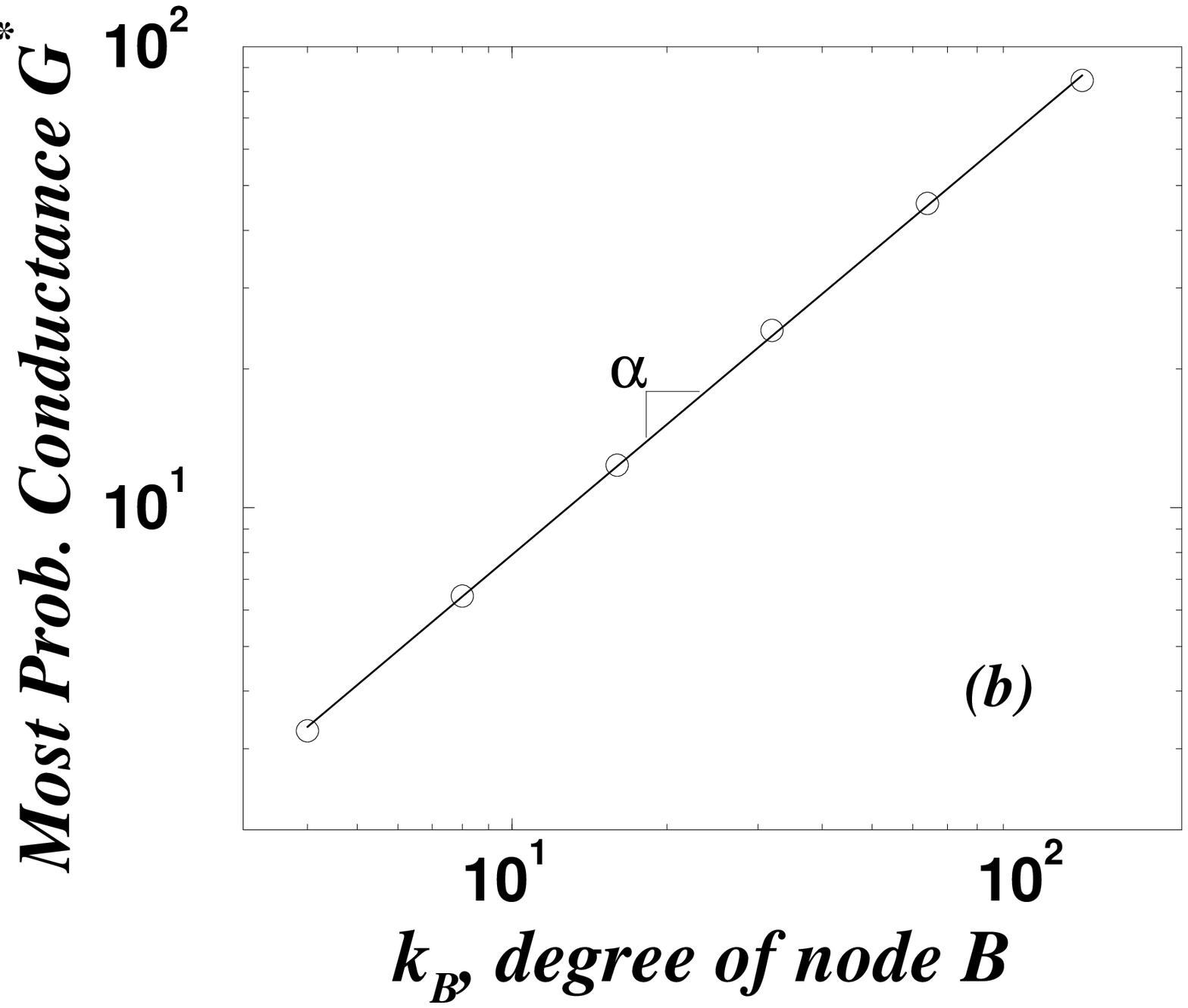,height=6cm,width=6cm}
\end{center}
\caption{(a) The pdf $\Phi_{\rm SF}(G|k_A,k_B)$ vs. $G$ for $N=8000$,
     $\lambda=2.5$ and $k_A=750$ ($k_A$ is close to the typical maximum
     degree $k_{\rm max}=800$ for $N=8000$). (b) Most probable values
     $G^*$, estimated from the maxima of the distributions in
     Fig.~\ref{PGka750_kb4-128_N8000_ab7_lamb2.5}(a), as a function of
     the degree $k_B$. The data support a power law behavior $G^*\sim
     k_B^{\alpha}$ with $\alpha=0.96\pm 0.05$.}
\label{PGka750_kb4-128_N8000_ab7_lamb2.5}
\end{figure}

We test this hypothesis by large scale computer simulations of the
conditional pdf $\Phi_{\rm SF}(G|k_A,k_B)$ for specific values of the
entering and exiting node degrees $k_A$ and $k_B$.  Consider first
$k_B\ll k_A$, and the effect of increasing $k_B$, with $k_A$ fixed.  We
find that $\Phi_{\rm SF}(G|k_A,k_B)$ is narrowly peaked
(Fig.~\ref{PGka750_kb4-128_N8000_ab7_lamb2.5}(a)) so that it is well
characterized by $G^*$, the value of $G$ when $\Phi_{\rm SF}$ is a
maximum.  We find similar results for Erd\H{o}s-R\'{e}nyi networks.
Further, for increasing $k_B$, we find
[Fig.~\ref{PGka750_kb4-128_N8000_ab7_lamb2.5}(b)] $G^*$ increases as
$G^*\sim k_B^{\alpha}$, with $\alpha=0.96\pm 0.05$ consistent with the
possibility that as $N\rightarrow\infty$, $\alpha=1$ which we assume
henceforth.

For the case of $k_B\gtrsim k_A$, $G^*$ increases less fast than
$k_B$, as can be seen in Fig.~\ref{G_over_kb_vs_ka_over_kb_combi_N8000_lamb2.5_ab7_m2}
where we plot $G^*/k_B$ against the scaled degree $x\equiv k_A/k_B$. The
collapse of $G^*/k_B$ for different values of
$k_A$ and $k_B$ indicates that $G^*$ scales as
\begin{equation}
G^*\sim k_Bf\left(\frac{k_A}{k_B}\right).
\label{G_max_scaled}
\end{equation}
Below we study the possible origin of this function.

\section{Transport backbone picture}
\label{theory}

\begin{figure}[t]
\begin{center}
\epsfig{file=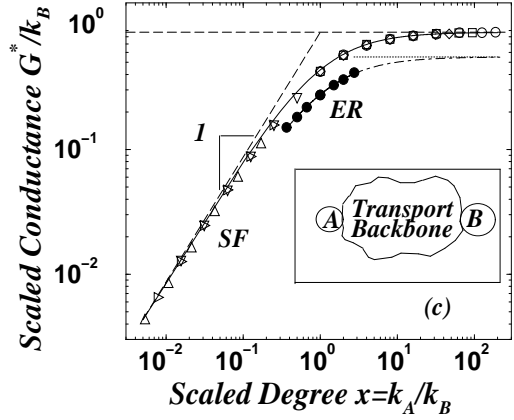,height=6cm,width=6cm}
\end{center}
\caption{ Scaled most probable
     conductance $G^*/k_B$ vs. scaled degree $x\equiv k_A/k_B$ for
     system size $N=8000$ and $\lambda=2.5$, for several values of $k_A$
     and $k_B$: $\Box$ ($k_A=8$, $8\le k_B\le 750$), $\diamondsuit$
     ($k_A=16$, $16\le k_B\le 750$), $\bigtriangleup$ ($k_A=750$, $4\le
     k_B\le 128$), $\bigcirc$ ($k_B=4$, $4\le k_A\le 750$),
     $\bigtriangledown$ ($k_B=256$, $256\le k_A\le 750$), and
     $\triangleright$ ($k_B=500$, $4\le k_A\le 128$).  The curve
     crossing the symbols is the predicted function
     $G^*/k_B=f(x)=cx/(1+x)$ obtained from Eq.~(\ref{e4b}).  We also
     show $G^*/k_B$ vs. scaled degree $x\equiv k_A/k_B$ for
     Erd\H{o}s-R\'{e}nyi networks with $\overline{k}=2.92$, $4\le k_A\le
     11$ and $k_B=4$ (symbol $\bullet$). The curve crossing the symbols
     represents the theoretical result according to Eq.~(\ref{e4b}), and
     an extension of this line to represent the limiting value of
     $G^*/k_B$ (dotted-dashed line).  The probability of observing $k_A>11$
     is extremely small in Erd\H{o}s-R\'{e}nyi networks, and thus we are
     unable to obtain significant statistics. The scaling function $f(x)$,
     as seen here, exhibits a crossover from a linear behavior to the
     constant $c$ ($c=0.87\pm0.02$ for scale-free networks, horizontal
     dashed line, and $c=0.55\pm0.01$ for Erd\H{o}s-R\'{e}nyi, dotted
     line).  The inset shows a schematic of the ``transport backbone''
     picture, where the circles labeled $A$ and $B$ denote nodes $A$ and
     $B$ and their associated links which do not belong to the
     ``transport backbone''.}
\label{G_over_kb_vs_ka_over_kb_combi_N8000_lamb2.5_ab7_m2}
\end{figure}

The behavior of the scaling function $f(x)$ can be interpreted using the
following simplified ``transport backbone'' picture
[Fig.~\ref{G_over_kb_vs_ka_over_kb_combi_N8000_lamb2.5_ab7_m2} inset], for which the
effective conductance $G$ between nodes $A$ and $B$ satisfies
\begin{equation}
{1\over G}={1\over G_A}+{1\over G_{tb}}+{1\over G_B},
\label{e4a}
\end{equation}
where $1/G_{tb}$ is the resistance of the ``transport backbone'' while
$1/G_A$ (and $1/G_B$) are the resistances of the set of links near node
$A$ (and node $B$) not belonging to the ``transport backbone''. It is
plausible that $G_A$ is linear in $k_A$, so we can write
$G_A=ck_A$. Since node $B$ is equivalent to node $A$, we expect
$G_B=ck_B$. Hence
\begin{equation}
G= \frac{1}{1/ck_A +1/ck_B+1/G_{tb}}
=k_B\frac{ck_A/k_B}{1+k_A/k_B+ck_A/G_{tb}},
\label{Gh}
\end{equation}
so the scaling function defined in Eq.~(\ref{G_max_scaled}) is
\begin{equation}
f(x)={cx\over 1+x+ck_A/G_{tb}}\approx{cx\over1+x}.
\label{e4b}
\end{equation}
The second equality follows if there are many parallel paths on the
``transport backbone'' so that $1/G_{tb}\ll 1/ck_A$ \cite{text1}.  The
prediction (\ref{e4b}) is plotted in
Fig.~\ref{G_over_kb_vs_ka_over_kb_combi_N8000_lamb2.5_ab7_m2}
for both scale-free and Erd\H{o}s-R\'{e}nyi networks
and the agreement with
the simulations supports the approximate validity of the transport
backbone picture of conductance in scale-free and Erd\H{o}s-R\'{e}nyi
networks.

The agreement of (\ref{e4b}) with simulations has a striking
implication: the conductance of a scale-free and Erd\H{o}s-R\'{e}nyi networks
depends on only one
quantity $c$. Further, since the distribution of
Fig.~\ref{PGka750_kb4-128_N8000_ab7_lamb2.5}(a) is sharply peaked, a
single measurement of $G$ for any values of the degrees $k_A$ and $k_B$
of the entrance and exit nodes suffices to determine $G^\ast$, which
then determines $c$ and hence through Eq.~(\ref{e4b}) the conductance
for all values of $k_A$ and $k_B$.

\begin{figure}[t]
\begin{center}
\epsfig{file=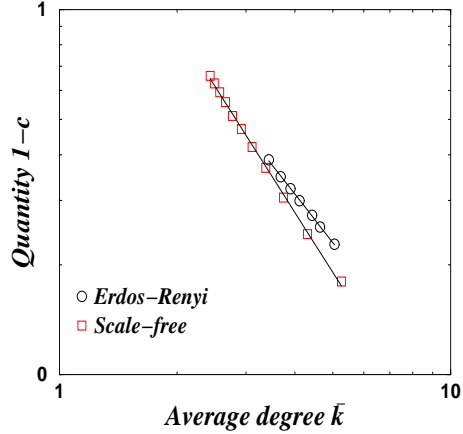,height=6cm,width=6cm}
\end{center}
\caption{Parameter $1-c$ vs. $\overline{k}$ for scale-free and
Erd\H{o}s-R\'{e}nyi networks with $N=8000$. The scale-free
networks display a power-law decay with exponent $-1.69\pm 0.02$,
whereas the Erd\H{o}s-R\'{e}nyi networks exhibit a decay exponent
of $-1.37\pm 0.02$.} \label{c_vs_lambda2.5-4.5_AND_er_8000}
\end{figure}

With regards to quantity $c$, first note it should grow, up to its upper
limit $1$, as the number of connections increases. For instance, a complete
graph has conductance $N/2$ which, if compared to Eq.~(\ref{Gh}), indicates
that indeed $c\rightarrow 1$. This suggests testing $c$ as a function
of the average degree $\overline{k}$. In
Fig.~\ref{c_vs_lambda2.5-4.5_AND_er_8000} we
present results for both scale-free and Erd\H{o}s-R\'{e}nyi networks.
The most important feature is that there seems to be a power-law
decay of $1-c$ with respect to $\overline{k}$. We find that
the dependence is of the form $1-c\sim \overline{k}^q$, with $q=-1.37\pm 0.02$
for Erd\H{o}s-R\'{e}nyi and $q=-1.69\pm 0.02$ for scale-free.
Also, we observe that
$c$ for Erd\H{o}s-R\'{e}nyi networks, at least in the region of $\overline{k}$
studied, is lower than for scale-free networks.
As $\overline{k}$ increases, transport on scale-free networks becomes
increasingly better than in Erd\H{o}s-R\'{e}nyi networks, because
$c$ is closer to one for the same $\overline{k}$.

\begin{figure}[t]
\begin{center}
\epsfig{file=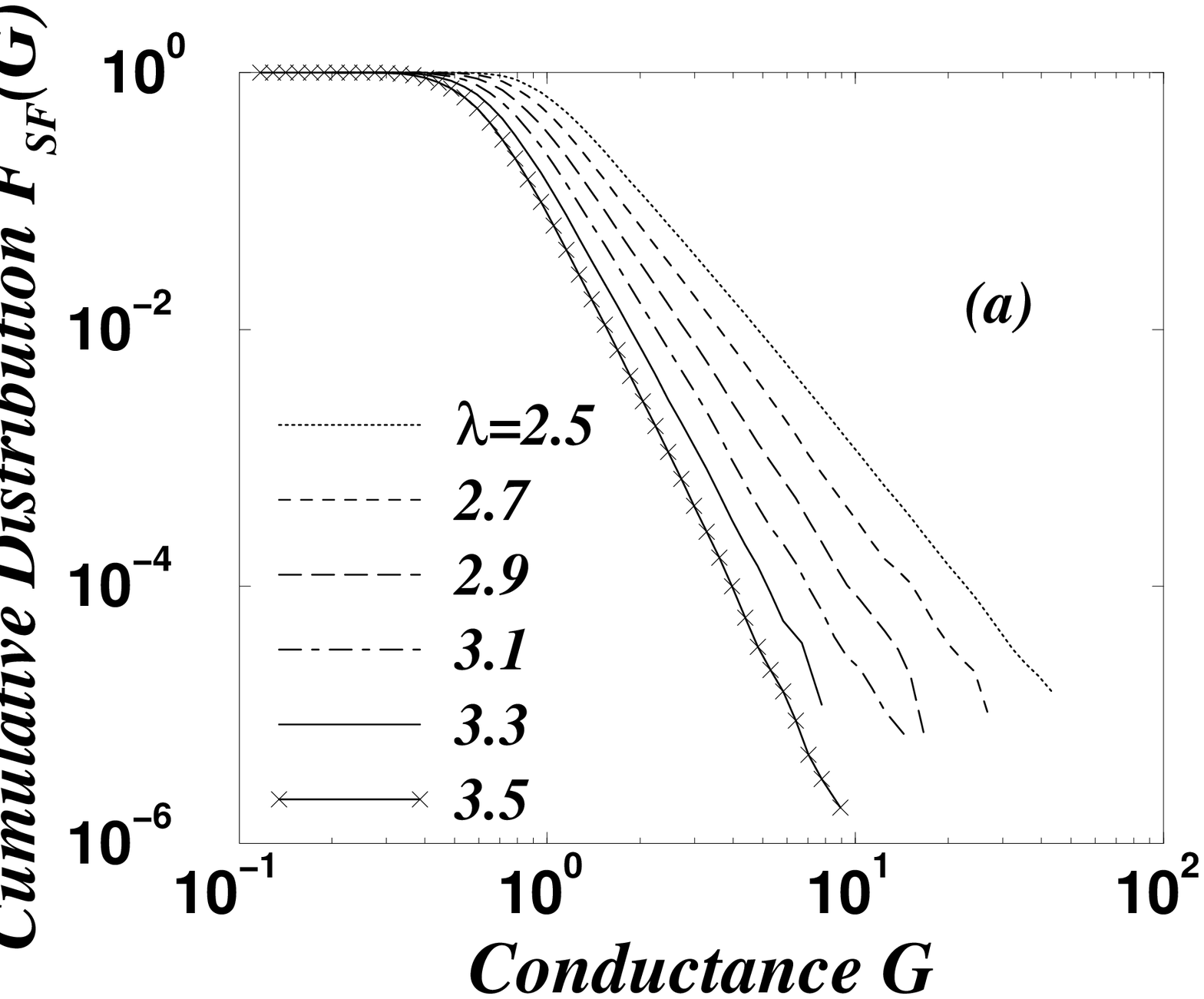,height=6.6cm,width=6cm}
\epsfig{file=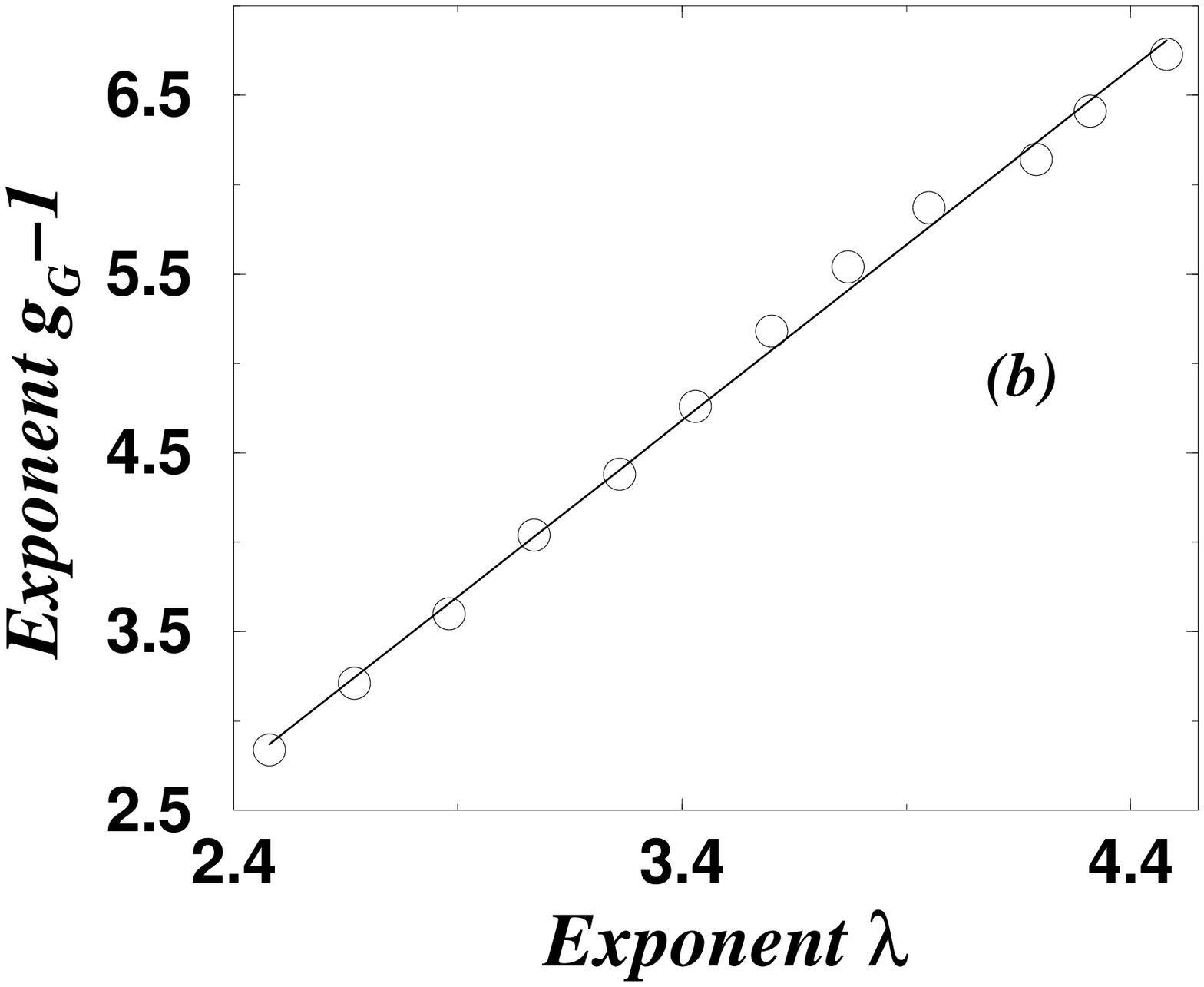,height=5.8cm,width=6cm}
\end{center}
\caption{(a) Simulation results for the cumulative distribution $F_{\rm
     SF}(G)$ for $\lambda$ between 2.5 and 3.5, consistent with the
     power law $F_{\rm SF}\sim G^{-(g_G-1)}$ (cf. Eq.~(\ref{PhiG})),
     showing the progressive change of the slope $g_G-1$. (b) The
     exponent $g_G-1$ from simulations (circles) with $2.5<\lambda<4.5$;
     shown also is a least
     square fit $g_G -1=(1.97\pm 0.04)\lambda -(2.01\pm 0.13)$,
     consistent with the predicted expression $g_G-1=2\lambda -2$
     [cf. Eq.~(\ref{PhiG})].}
\label{FG_G_lamb2.5-3.5_8000}
\end{figure}

Within this ``transport backbone'' picture, we can analytically
calculate $F_{\rm SF}(G)$.
The key insight necessary for this calculation is that $G^*\sim k_B$, when
$k_B\leq k_A$, and we assume that $G\sim k_B$ is also
valid given the narrow shape of $\Phi_{\rm SF}(G|k_A,k_B)$.
This implies that the probability of observing
conductance $G$ is related to $k_B$ through
$\Phi_{\rm SF}(G)dG\sim M(k_B)dk_B$, where $M(k_B)$ is the probability that,
when to nodes $A$ and $B$ are chosen at random, $k_B$ is the minimum
degree. This can be calculated analytically through
\begin{equation}
M(k_B)\sim P(k_B)\int^{k_{\rm max}}_{k_B} P(k_A) dk_A
\end{equation}
Performing the integration 
we obtain for $G<G_{\rm max}$
\begin{equation}
\Phi_{\rm SF}(G)\sim G^{-g_G} \qquad [g_G=2\lambda -1].
\label{PhiG}
\end{equation}
Hence, for $F_{\rm SF}(G)$, we have
$F_{\rm SF}(G)\sim G^{-(2\lambda -2)}$.
To test this prediction, we perform simulations for scale-free networks
and calculate the values of $g_G-1$ from the slope of a log-log plot of
the cumulative distribution $F_{\rm SF}(G)$.  From
Fig.~\ref{FG_G_lamb2.5-3.5_8000}(b) we find that
\begin{equation}
g_G -1=(1.97\pm 0.04)\lambda -(2.01\pm 0.13).
\label{g_G_measured}
\end{equation}
Thus, the measured slopes are consistent with the theoretical values
predicted by Eq.~(\ref{PhiG}) \cite{fn_lambda}.

\begin{figure}[t]
\begin{center}
\epsfig{file=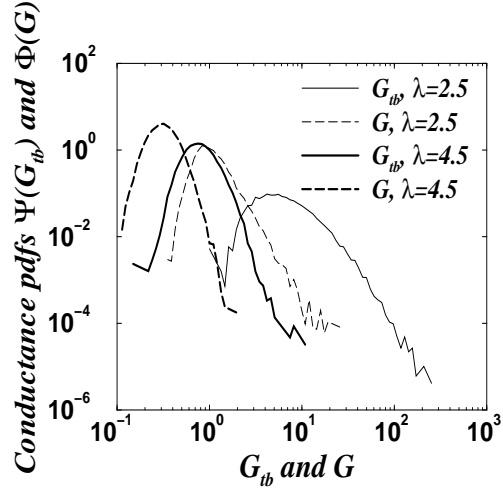,height=6cm,width=6cm}
\end{center}
\caption{Comparison of pdf $\Psi(G_{tb})$ and $\Phi(G)$ for networks of
$N=8000$ for two values of $\lambda$.}
\label{Gtb_and_G_fig_lamb2.5-4.5}
\end{figure}

The transport backbone conductance $G_{tb}$ of scale-free networks
can also be studied through its pdf $\Psi_{\rm SF}$
(see Fig.~\ref{Gtb_and_G_fig_lamb2.5-4.5}).
To determine $G_{tb}$, we consider the contribution to the conductance of
the part of the network with paths between $A$ and $B$,
excluding the contributions from the vicinities of nodes $A$ and $B$, which
are determined by the quantity $c$.
The most relevant feature in Fig.~\ref{Gtb_and_G_fig_lamb2.5-4.5} is that,
for a given $\lambda$ value, both $\Psi_{\rm SF}$
and $\Phi(G)$ have equal decay exponents, suggesting that
they are also related to $\lambda$ as Eq.~(\ref{g_G_measured}).
Figure~\ref{Gtb_and_G_fig_lamb2.5-4.5} also shows that the values of $G_{tb}$
are significantly larger than $G$.

\section{Number of link-independent paths: transport without friction}
\label{frictionless}

In many systems, it is the nature of the transport process that
the particles flowing through the network links experience no
friction. For example, this is the case in an electrical system
made of super-conductors \cite{kirk}, or in the case of water
flow along pipes, if frictional effects are minor. Other examples
are flow of cars along traffic routes, and perhaps most important,
the transport of information in communication networks. Common to
all these processes is that, the quality of the transport is
determined by the number of link-independent paths leading from the source
to the destination (and the capacity of each path), and not by the
length of each path (as is the case for simple electrical
conductance). In this section, we focus on non-weighted networks, and
define the conductance, as
the number of link-independent paths between a given source and
destination $A$ and $B$. We name this transport process as the
\emph{max-flow model}, and denote the conductance as $G_{\rm MF}$.
Fast algorithms for solving the max-flow problem, given a network
and a pair $(A,B)$ are well known within the computer science
community \cite{maxflow}. We apply those methods to random
scale-free and Erd\H{o}s-R\'{e}nyi networks, and observe
similarities and differences from the electrical conductance
transport model. Max-flow analysis has been applied recently for
complex networks in general \cite{Lee,Wu}, and for the Internet in
particular \cite{internet}, where it was used as a significant
tool in the structural analysis of the underlying network.

\begin{figure}[t]
\begin{center}
\epsfig{file=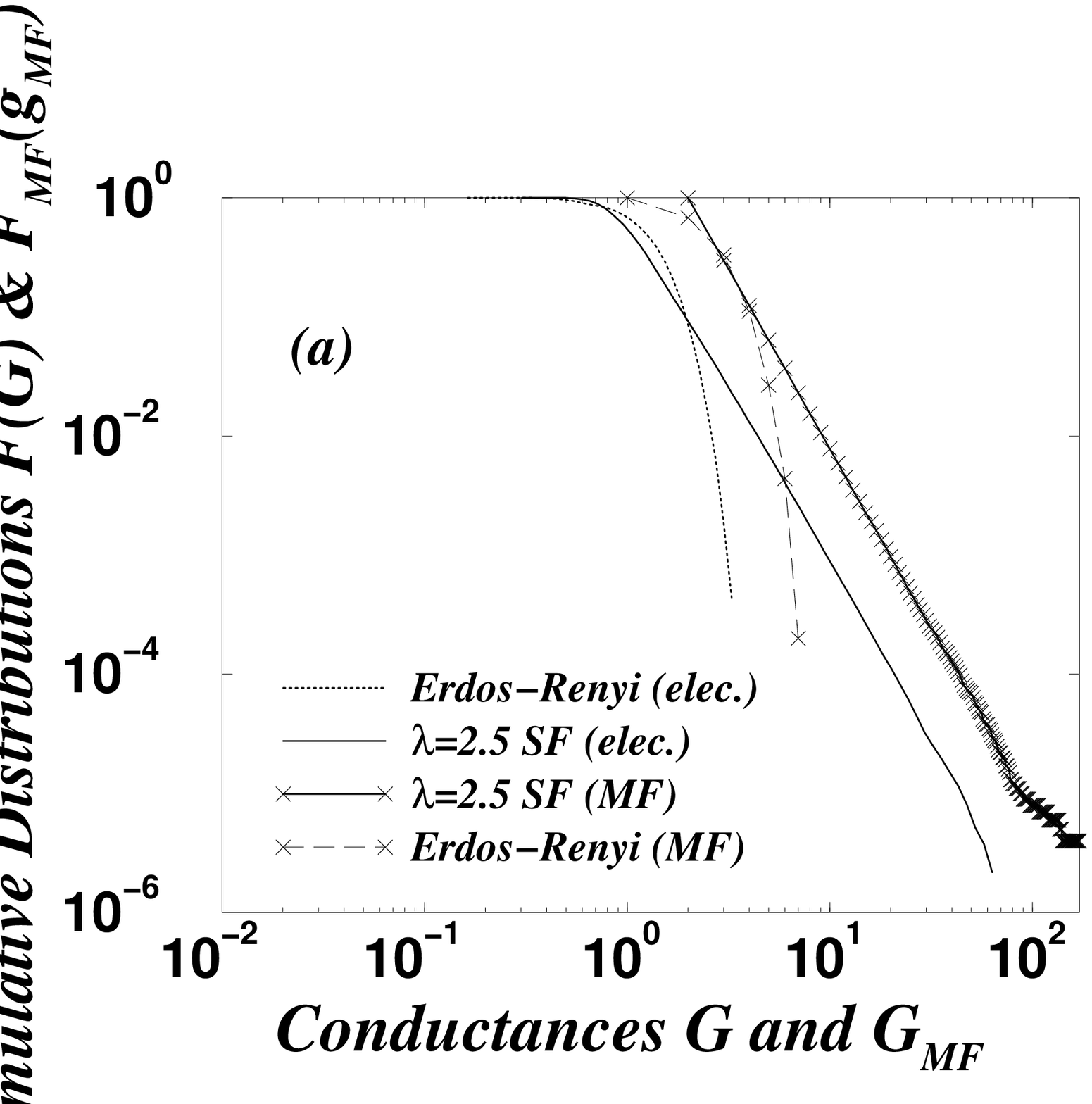,height=6cm,width=6cm}
\epsfig{file=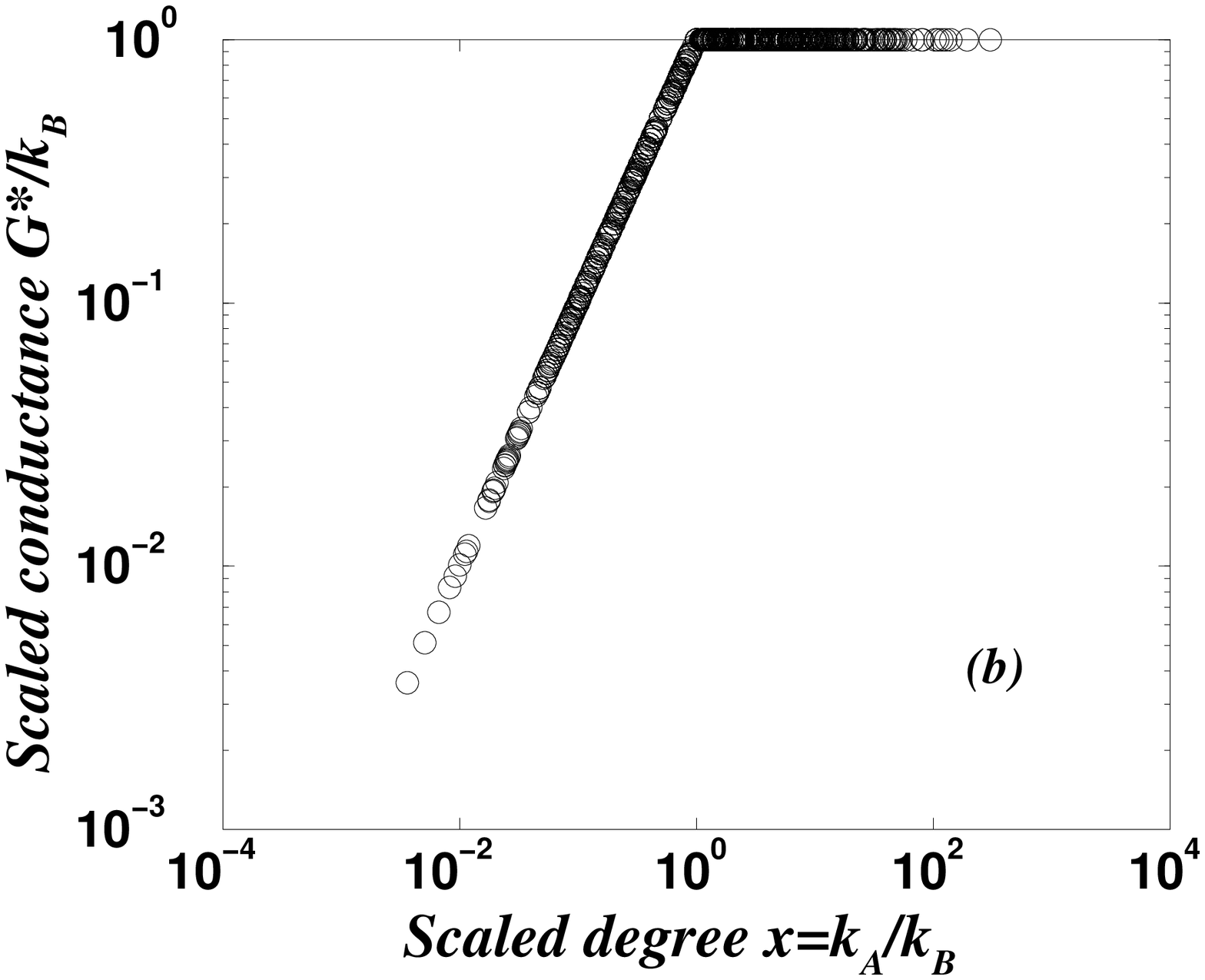,height=6cm,width=6.8cm}
\end{center}
\caption{(a) Cumulative distribution of link-independent paths
(conductance) $F_{\rm MF}(G_{\rm MF})$ vs. $G_{\rm MF}$ compared
with the electrical conductance distributions taken from
Fig.~\ref{FG_lamb2.5-3.3-ER_N8000}. We see that the scaling is
indeed the same for both models, but the proportionality constant
of $F_{\rm MF}(G_{\rm MF})$ vs. $G_{\rm MF}$ is larger for the
frictionless problem. (b) Scaled most probable number of
independent paths $G_{\rm MF}^*/k_B$ as a function of the scaled
degree $k_A/k_B$ for scale-free networks of $N=8000$,
$\lambda=2.5$ and $k_{min}=2$. The behavior is sharp, and shows
how $G_{\rm MF}^*$ is a function of only the minimum $k$.}
\label{flow_cumulative_distribution}
\end{figure}

We find, that in the max-flow model, just as in the electrical
conductance case, scale-free networks exhibit a power-law decay of
the distribution of conductances with the same exponent 
(and thus very high conductance values are
possible), while in Erd\H{o}s-R\'{e}nyi networks, the conductance
decays exponentially (Fig.~\ref{flow_cumulative_distribution}(a)).
In order to better understand this behavior, we plot the
scaled-flow $G_{\rm MF}^*/k_B$ as a function of the scaled-degree
$x \equiv k_A/k_B$ (Fig.~\ref{flow_cumulative_distribution}(b)). It
can be seen that the transition at $x=1$ is sharp. For all $x < 1$
($k_A < k_B$), $G_{\rm MF}^*=x$ (or $G_{\rm MF}^* = k_A$), while
for $x > 1$ ($k_B < k_A$), $G_{\rm MF}^*=1$ (or $G_{\rm MF}^* =
k_B$). In other words, the conductance simply equals the minimum
of the degrees of $A$ and $B$. In the symbols of Eq. (\ref{Gh}),
this also implies that $c \rightarrow 1$; i.e. scale-free networks
are optimal for transport in the max-flow sense. The derivation
leading to Eq. (\ref{PhiG}) becomes then exact, so that the
distribution of conductances is given again by $\Phi_{\rm
MF,SF}(G_{\rm MF}) \sim G_{\rm MF}^{-(2\lambda-1)}$.

\begin{figure}
\begin{center}
\epsfig{file=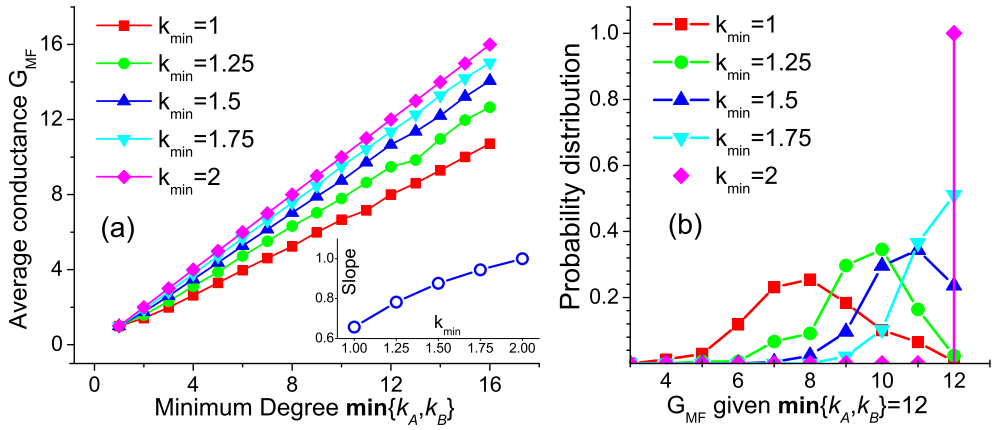,height=7cm,width=14.5cm}
\end{center}
\caption{(a) Average conductance $\overline{G}_{\rm MF}$ vs.
minimum degree of the source and sink $A$ and $B$ for different
values of $k_{min}$, the minimum degree in the network. All curves
show the behavior $\overline{G}_{\rm MF} \propto k$, as the
proportionality coefficient gradually increases (see inset), until
eventually becomes 1 as $k_{min}$ approaches 2. (b) The same
concept is illustrated by plotting the probability to find a
specific conductance $G_{\rm MF}$} when the minimum degree is 12,
for few values of $k_{min}$. \label{flow_kmin}
\end{figure}

This picture of the transport is seen when the minimum degree in
the network is $k_{min} = 2$. When the minimum degree is allowed to
take values in the range between 1 and 2~\cite{fn_kmin}, we find
that $G_{\rm MF} \propto \min \{k_A,k_B\}$, but the two quantities
are no longer equal. This reflects the fact that as the minimum
network degree is lowered, the network becomes more dilute, such
that two paths starting at the source might intersect at some link
inside the backbone. In other words, the conductance of the
backbone is still high, but no longer infinite.
This is illustrated in Fig.~\ref{flow_kmin}(a), where we plot the
average conductance $\overline{G}_{\rm MF}$ vs. the minimum degree
of the source and sink $\min \{k_A,k_B\}$, and find that while the
relation between the two variables is linear, the slope is not
necessarily 1. Nevertheless, as $k_{min}$ approaches 2, the slope
becomes 1, which indicates that a sufficient condition for the
network to have infinite backbone conductivity is $k_{min} \geq
2$. This is illustrated again in Fig.~\ref{flow_kmin}(b), where
the distribution of conductance values $G_{\rm MF}$ for fixed
$\min \{k_A,k_B\}$ is plotted.

\begin{figure}
\begin{center}
\epsfig{file=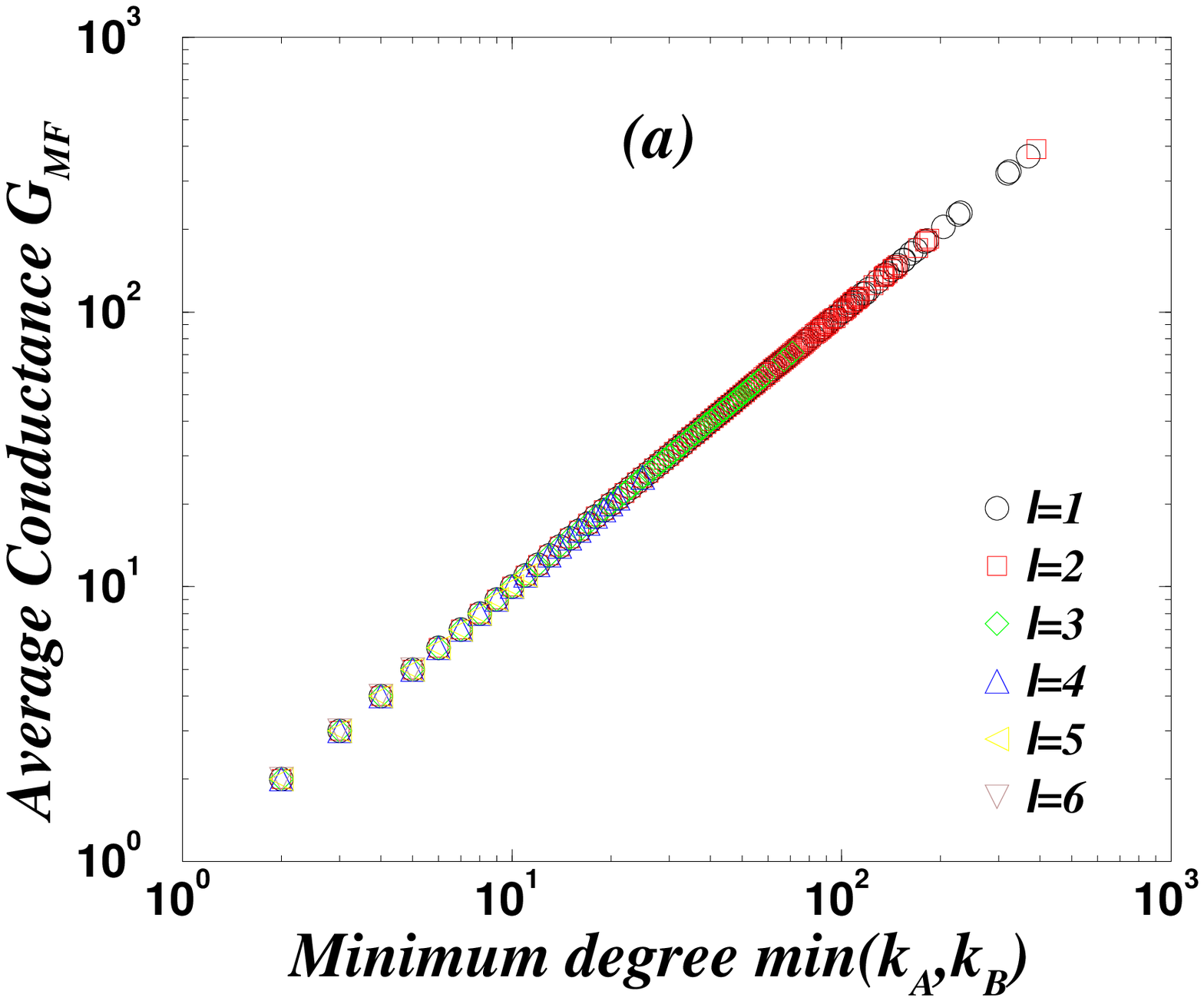,height=6cm,width=6cm}
\epsfig{file=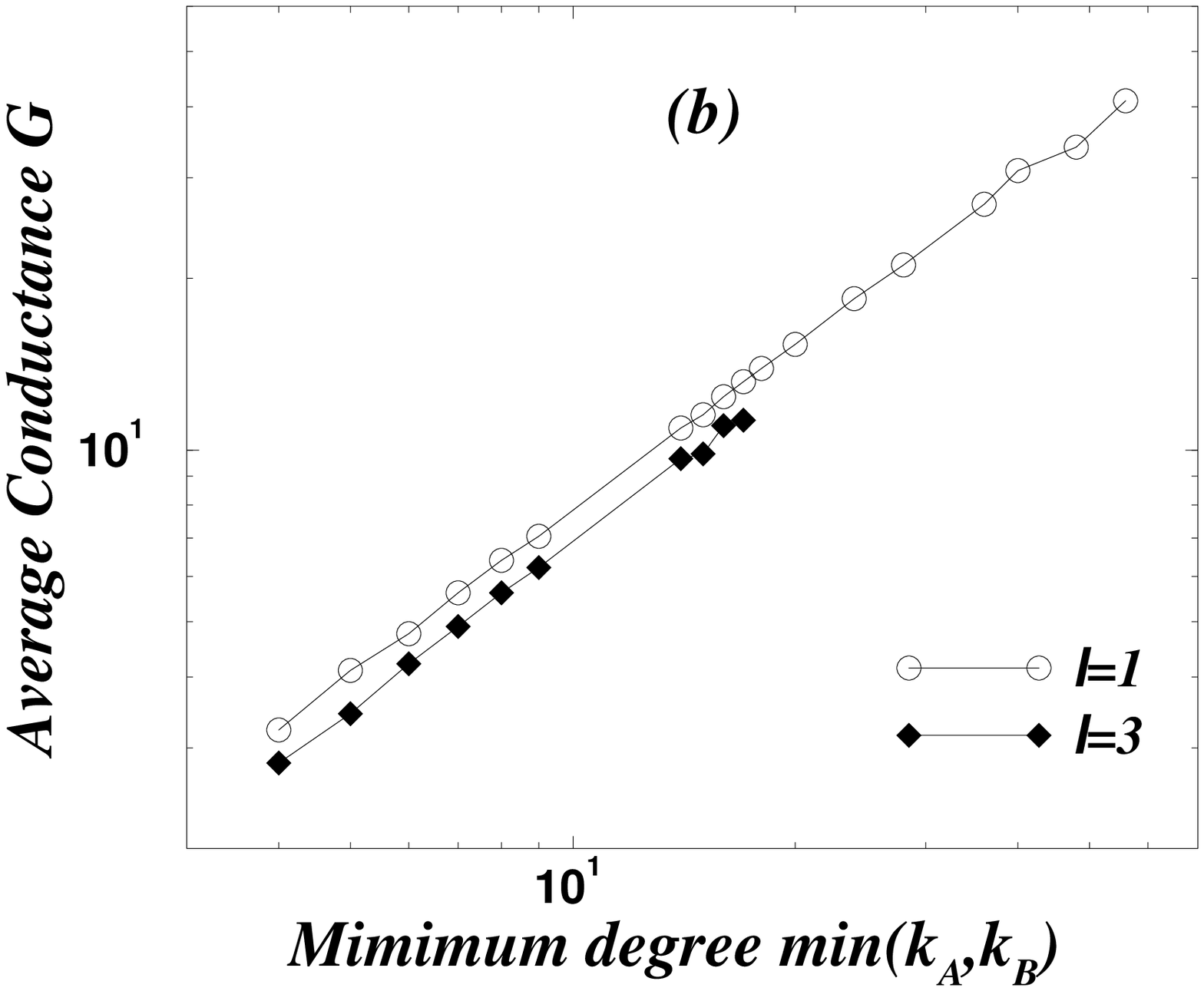,height=6cm,width=6cm}
\end{center}
\caption{(a) Average conductance $\overline{G}_{\rm MF}$ vs.
minimum degree $\min(k_A,k_B)$ of the source and sink $A$ and $B$ for different
values of the shortest distance $\ell_{AB}$. The relation is
independent of $\ell_{AB}$ indicating the independence of
$\overline{G}_{\rm MF}$ on the distance. The network has $N=8000$,
$\lambda=2.5$, $k_{min}=2$. (b) Average conductance $\overline{G}$
vs. minimum degree $\min(k_A,k_B)$ of the source and sink $A$ and $B$ for
different values of distance $\ell_{AB}$. The independence of
$\overline{G}$ with respect to $\ell_{AB}$ breaks down and, as
$\ell_{AB}$ increases, $\overline{G}$ decreases. Once again,
$N=8000$ and $\lambda=2.5$, but the average has been performed for
various $k_B<k_A$ and $k_A=750$.} \label{flow_distance}
\end{figure}

We have so far observed that the max-flow model is quite similar
to electrical conductance, by means of having a finite possibility
of finding very high values of conductance. Also, the fact that
the minimum degree plays a dominant role in the number of
link-independent paths makes the scaling behavior of the
electrical and frictionless problems similar. Only when the
conductances are studied as a function of distance, some
differences between the electrical and frictionless cases begin to
emerge. In Fig,~\ref{flow_distance}(a), we plot the dependence of
the average conductance $\overline{G}_{\rm MF}$ with respect to
the minimum degree $\min(k_A,k_B)$ of the source and sink, for different values of
the shortest distance $\ell_{AB}$ between $A$ and $B$, and find
that $\overline{G}_{\rm MF}$ is independent of $\ell_{AB}$ as the
curves for different $\ell_{AB}$ overlap. This result is a
consequence of the frictionless character of the max-flow problem.
However, when we consider the electrical case, this independence
disappears. This is illustrated in Fig,~\ref{flow_distance}(b),
where $\overline{G}$ is also plotted against the minimum degree
$\min(k_A,k_B)$,
but in this case, curves with different $\ell_{AB}$ no longer
overlap. From the plot we find that $\overline{G}$ decreases as
the distance increases. This is explained using the observation of
\cite{Holyst}, that the average shortest distance between the
source and the sink is inversely proportional to the (logarithm of
the product) of their degrees. Thus, on average, shorter distances
are attributed to higher degrees, which in turn are connected by
larger conductance. Finally, it is interesting to note that the
dependence of $\overline{G}$ with respect to $\ell_{AB}$ is slower
than linear.

\begin{figure}
\begin{center}
\epsfig{file=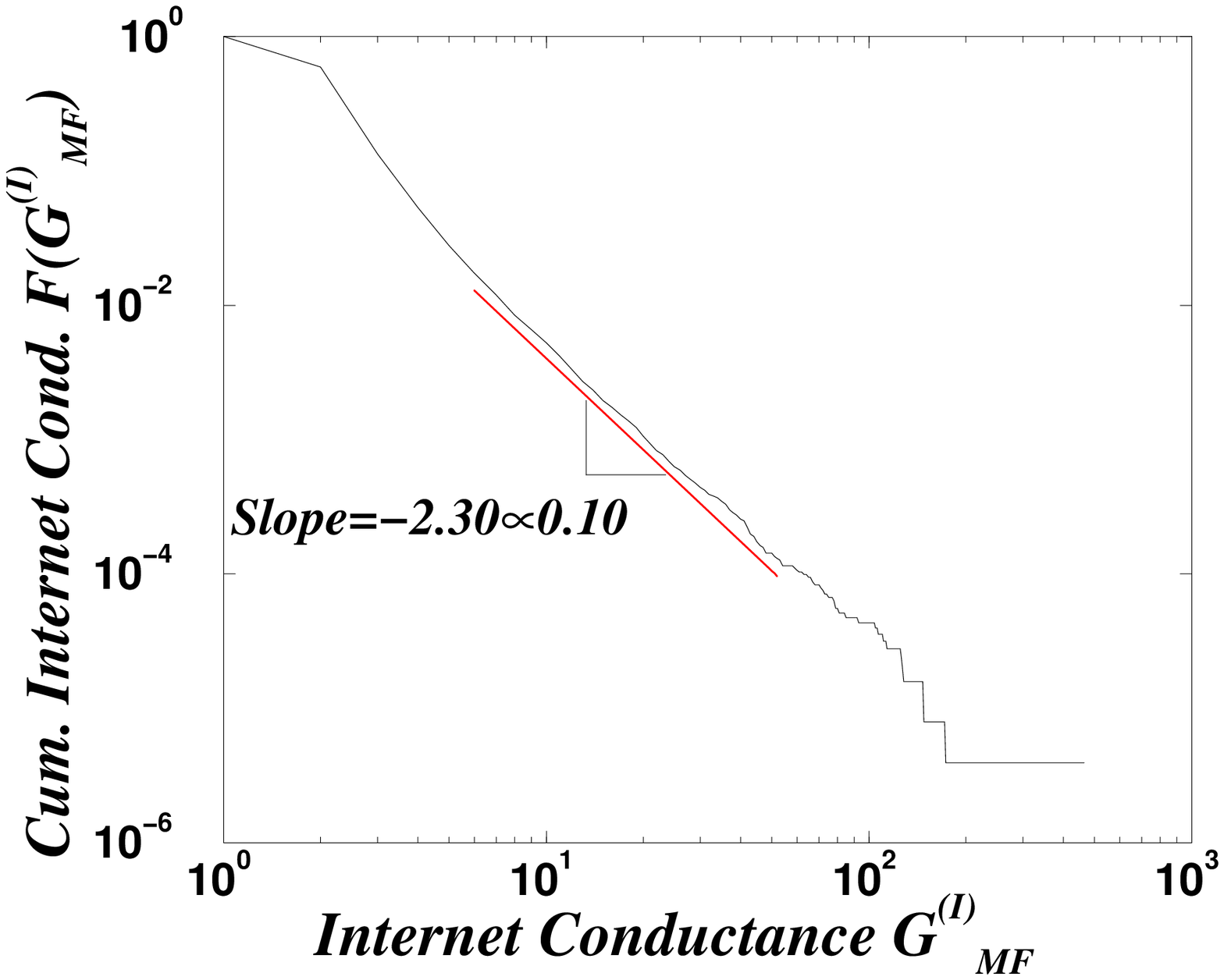,height=6cm,width=6cm}
\end{center}
\caption{Cumulative distribution $F(G^{(I)}_{\rm MF})$ of
$G^{(I)}_{\rm MF}$ for the Internet. This data set is consistent
with the scale-free structure that has been predicted for the
Internet (see text).} 
\label{flow_internet}
\end{figure}

In order to test the validity of our results in real networks, we measured the
conductance $G^{(I)}_{\rm MF}$ on the most up to date map of the
Autonomous Systems (AS) level of the Internet
structure~\cite{Shavitt}. From Fig.~\ref{flow_internet} we find that 
the slope of the plot, which corresponds to $g_G-1$ from Eq.~(\ref{PhiG}), 
is approximately 2.3, 
implying that $\lambda\approx 2.15\pm0.05$. This value of
$\lambda$ is in good agreement with 
the value of the degree distribution exponent for
the Internet observed in~\cite{Shavitt}.

\section{Summary}
\label{summary}

In summary, we find that the conductance of scale-free networks is
highly heterogeneous, and depends strongly on the degree of the
two nodes $A$ and $B$. Our results suggest that the transport
constants are also heterogeneous in these networks, and depend on
the degrees of the starting and ending nodes. We also find a
power-law tail for $\Phi_{SF}(G)$ and relate the tail exponent
$g_G$ to the exponent $\lambda$ of the degree distribution $P(k)$.
This power law behavior makes scale-free networks better for
transport. Our work is consistent with a simple physical picture
of how transport takes place in scale-free and Erd\H{o}s-R\'{e}nyi
networks. This, so called ``transport backbone'' picture consists
of the nodes $A$ and $B$ and their vicinities, and the rest of the
network, which constitutes the transport backbone.  Because of the
great number of parallel paths contained in the transport
backbone, transport takes place inside with very small resistance,
and therefore the dominating effect of resistance comes from the
vicinity of the node ($A$ or $B$) with the smallest degree. This
scenario appears to be valid for both the electrical and frictionless models,
as clearly indicated by the similarity in the results. The
quantity $c$, which characterizes transport for a complex network
exhibits a behavior of the form $1-\overline{k}^q$ for both
scale-free and Erd\H{o}s-R\'{e}nyi networks in the electrical
model, and in the frictionless model $c=1$ in most cases. We
observe that as $\overline{k}$ increases, scale-free networks
become progressively better than Erd\H{o}s-R\'{e}nyi networks in
electrical transport.

Finally, we point out that our study can be extended further.
For instance, it has been found recently that many real-world
scale-free networks possess fractal properties
\cite{song-makse-havlin}. However, random scale-free and
Erd\H{o}s-R\'{e}nyi networks, which are the subject of this study,
do not display fractality. Since fractal substrates also lead to
anomalous transport
\cite{havlin-ben-avraham,ben-avraham-havlin,bunde-havlin}, it
would be interesting to explore the effect of fractality on
transport and conductance in fractal networks. This case is
expected to have anomalous effects due to both the heterogeneity
of the degree distribution and to the fractality of the network.
Furthermore, the effect on conductivity and transport of the
correlation between distance of two nodes and their degree
\cite{Holyst} should be further investigated.

%
\section{Acknowledgements}
We thank the Office of Naval Research, the Israel Science Foundation,
the European NEST project DYSONET, and the Israel Internet Association
for financial support, and
L. Braunstein, R. Cohen, E. Perlsman, G. Paul, S. Sreenivasan,
T. Tanizawa, and Z. Wu for discussions.



\newpage

\newpage


\begin{thebibliography}{00}
\bibitem{netwcomment} In principle, a node can have a degree up to
$N-1$, connecting to all other nodes of the network.  The results
presented here correspond to networks with upper cutoff $k_{\rm
max}=k_{\rm min}N^{1/(\lambda-1)}$ imposed.  We also studied networks
for which $k_{\rm max}$ is not imposed, and found no significant
differences in the pdf $\Phi_{\rm SF}(G)$.
\bibitem{dyn-topology} The dynamical properties we study are
related to transport on networks and differ from those which treat the
network topology itself as evolving in time \cite{scale-Barabasi,dyn-network}.
\bibitem{fn1} We performed simulations
with the node copies randomly matched, and also
matched in order of degree
from highest to lowest and obtained similar results.
\bibitem{text1}
Flux starts at node $A$, and it is controlled by the conductance of the
bonds in the vicinity of $A$. This flux passes into the ``transport
backbone'', which is comprised of many parallel paths and hence has a
high conductance.  Finally, flux ends at node $B$,
controlled by the conductance of the bonds in the vicinity of $B$
(see inset of Fig.~\ref{G_over_kb_vs_ka_over_kb_combi_N8000_lamb2.5_ab7_m2}).
This is similar to traffic around a major
freeway.  Most of the limitations to transport occur in getting to the
freeway (``node $A$'') and then after leaving it (``node $B$''),
but flow occurs easily on the freeway (``transport backbone'').
\bibitem{fn_lambda} The $\lambda$ values explored here are limited by
computer time considerations. 
Most real-world examples of scale-free networks exhibit exponents 
that satisfy $2\leq\lambda\leq 4$ and are thus within the range of values
we simulate. In the case of very large $\lambda$, and for $k_{\rm min}$
large enough, scale-free networks gradually become similar
to Erd\H{o}s-R\'{e}nyi networks, and Eq.~(\ref{Gh}) continues to hold,
but the power-law behavior for $\Phi(G)$ disappears. When $k_{\rm min}$ is
small (close to 1), large $\lambda$ induces a behavior close to percolation,
where Eq.~(\ref{Gh}) no longer holds due to the tree-like structure
of the connected components.
\bibitem{fn_kmin} In order to obtain a value of mimimum degree between 1 and 2,
realizations of networks are generated that have a probability of having mimimum
degree either 1 or 2. Statistically, one can do this in such a way as to control the
average value of the mimimum degree by controling the proportion of networks
created with mimimum degree 1 compared to those created with mimimum degree 2, as is
done here.
\end{thebibliography}
\end{document}